\documentclass[12pt]{article}
\usepackage{epsfig}
\usepackage{cite}

\begin{document}
\begin{titlepage}

\centerline{\large\bf Center vortex model for the infrared sector of SU(4)}
\vspace{0.2cm}
\centerline{\large\bf Yang-Mills theory: String tensions and deconfinement
transition}

\bigskip
\centerline{M.~Engelhardt\footnote{\tt email:\ engel@nmsu.edu} }
\vspace{0.2 true cm}
\centerline{\em Physics Department, New Mexico State University}
\centerline{\em Las Cruces, NM 88003, USA}

\abstract{A random vortex world-surface model for the infrared sector of
$SU(4)$ Yang-Mills theory is constructed, focusing on the confinement
properties and the behavior at the deconfinement phase transition.
Although the corresponding data from lattice Yang-Mills theory can
be reproduced, the model requires a more complex action and considerably
more tuning than the $SU(2)$ and $SU(3)$ cases studied previously. Its
predictive capabilities are accordingly reduced. This behavior has
a definite physical origin, which is elucidated in detail in the present
work. As the number of colors is raised in Yang-Mills theory, the
corresponding infrared effective vortex description cannot indefinitely
continue to rely on dynamics determined purely by vortex world-surface
characteristics; additional color structures present on the vortices
begin to play a role. As evidenced by the modeling effort reported here,
definite signatures of this behavior appear in the case of four colors.}

\vspace{1cm}



\end{titlepage}

\section{Introduction}
The vortex picture of the strong interaction vacuum was originally conceived
\cite{hooft,aharonov,cornold,mack,olesen} as an explanation for confinement.
More recent work 
\cite{jg1,jg2,jg3,tk1,tk2,tk3,cw1,cw2,cw3,int,per,rb,df1,df2,df3,m1,m2,m3}
has established it as a
comprehensive, and in many respects quantitative infrared effective model
of the vacuum, generating not only confinement, but also the spontaneous
breaking of chiral symmetry and the axial $U_A (1)$ anomaly. The evidence
for the vortex picture rests on two pillars. On the one hand, methods have
been developed which permit the identification and isolation of vortex
structures in the gauge configurations of lattice QCD \cite{jg1,jg2,df2}.
Based on these techniques, the phenomenology induced by the vortices
present in lattice gauge configurations can be studied. Highlights of the
results obtained in this manner include vortex dominance of the string
tension \cite{jg1,jg2,df2}, the absence of nonperturbative effects (string
tension, chiral condensate, topological susceptibility) when vortices are
removed from lattice gauge configurations \cite{df1,df2}, and a natural
explanation of the deconfining phase transition as a vortex percolation
transition \cite{per}.

On the other hand, an infrared effective random vortex world-surface model
has been developed \cite{m1,m2,m3,su3conf,su3bary,su3freee,su3proc}
which demonstrates that the main nonperturbative
features of the strong interaction vacuum can be described\footnote{In the
$SU(2)$ and $SU(3)$ models studied to date, adjustment of a single
dimensionless coupling constant suffices to arrive at a phenomenologically
viable model, a number of nonperturbative observables being quantitatively
predicted as a result. The $SU(4)$ case investigated here turns out to behave
less favorably in this respect, and the reasons behind this represent a
principal focus of the present work.} on the basis of a {\em weakly coupled}
vortex dynamics. This further buttresses the notion that vortices constitute
the relevant infrared gluonic degrees of freedom in that vacuum. Through its
reduced set of degrees of freedom and the correspondingly simplified dynamics,
the random vortex world-surface model has allowed the investigation of a
wider range of scenaria than has been accessible using the vortices extracted
from lattice gauge configurations discussed further above. In the random
vortex world-surface model, not only has the $SU(2)$ case been investigated
comprehensively \cite{m1,m2,m3}, but also the confining properties of the
$SU(3)$ case have been studied in detail \cite{su3conf,su3bary,su3freee},
and the present work extends the model to $SU(4)$ color.

\section{Motivation}
\label{motiv}
\subsection{Vortex color structure}
To clarify the motivation for the present investigation, it is useful to
begin by reviewing the different color structures which can occur on
center vortices, specifically in various Abelian formulations. As will be
discussed below, vortices are not only characterized by their location
in space-time, but possess considerable additional structure in color
space. In general, it cannot be excluded that this structure may
significantly influence vortex dynamics; besides geometrical world-surface
characteristics, the vortex effective action potentially may also include
a dependence on the aforementioned color structure. Investigation of the
possibility or even necessity of such a dependence is a principal focus
of the present investigation. Thus, a review of vortex color structure
is in order.

Center vortices are closed tubes of quantized chromomagnetic flux in
three spatial dimensions. Accordingly, they are described by
(thickened) world-surfaces in four-dimensional (Euclidean) space-time.
The quantization is defined by the property that a Wilson loop encircling
a vortex (more precisely, linked to the latter) acquires a factor
corresponding to a center element of the gauge group. In $SU(N)$ gauge
theory, there are thus $N-1$ possible vortex fluxes (the trivial unit
element corresponds to no flux being present). In particular, this entails
that vortices can branch for $N\geq 3$, as described in greater detail
below.

This description of vortices in terms of their space-time location and
their influence on Wilson loops is in some respects incomplete. While it
is adequate for the discussion of the confinement properties encoded in
Wilson loops, to fully describe the topological properties of vortices and
the related chiral symmetry breaking phenomena, it is necessary to take
into account the direction of the vortex field strength in color
space \cite{cw1,cw2,cw3,m2,m3,contvort,bruck}.
While this direction can be rotated by gauge transformations, certain
properties of it carry gauge-independent significance. For instance,
generic vortex world-surfaces are nonorientable; this implies that the
field strength cannot be globally proportional to a constant color vector.
Instead, the color direction of the field strength must vary, in any gauge.
While the detailed local space-time form of the variation forced by vortex
nonorientability is gauge-dependent, it contains invariant characteristics
which manifest themselves, e.g., in the topological charge.

Often, it is convenient to construct vortex configurations in an Abelian
gauge. In this case, variations of the field strength color vector are
compressed into lines on the vortex world-surfaces, leaving the color
vector constant elsewhere. At these lines, the field strength is
discontinuous in a way which corresponds to the presence of a source or
sink of magnetic flux, i.e., an Abelian magnetic monopole.
Thus, due to the nonorientability of vortex world-surfaces, monopoles
represent an intrinsic feature of vortex configurations in Abelian gauges.

From this more detailed perspective, in particular also the branching
points (in three-dimensional space) or lines (in four-dimensional
space-time) of vortices mentioned further above can be characterized
more thoroughly. Vortex branchings in general are associated with
nontrivial rearrangements in the vortex color structure; they cannot
be realized while keeping all vortex field strengths in the branching
region pointing into the same fixed color direction, as will be
discussed below.

To facilitate this discussion, (Abelian) vortex color direction will be
specified in the following by giving the (diagonal) chromomagnetic flux
matrix $Q$ which results when evaluating a closed line integral encircling
the vortex,
\begin{equation}
Q=\oint dx_{\mu } A_{\mu } \ .
\end{equation}
This eliminates the detailed space-time information on the vortex field
strength which is non-essential to the following discussion. In terms of
$Q$, the Wilson loop along the same path is 
\begin{equation}
W=\frac{1}{N} \mbox{ Tr } \exp (iQ) \ .
\end{equation}
Note that every vortex world-surface can be associated with two possible
orientations, which yield opposite signs for $Q$. Conversely, of course,
inverting the integration direction in the line integral defining $Q$
inverts the sign of the latter. In order to preserve rotational symmetry
in the following, if a given description allows for vortices associated
with the color matrix $Q$, it will also include those associated with
the color matrix $-Q$.

\subsection{Minimal Abelian construction}
\label{minconst}
The color structure of vortex branchings can now be viewed in a variety of
ways. One possible picture results if one chooses a minimal set of definite
(Cartan) color directions to describe the $N-1$ types of vortex flux
permitted by the gauge group. Under this very restrictive condition,
vortex branching is tied to the presence of magnetic monopoles. Consider
the $SU(3)$ gauge group as an example. The two nontrivial center elements
are
\begin{equation}
\exp ( 2\pi i/3) \ , \ \ \ \exp ( -2\pi i/3) \ .
\end{equation}
These are the phases which can arise in a Wilson loop due to linking with
vortices. Since these two phases are complex conjugates of one another,
they can actually be generated by the two possible orientations of one
single type of vortex. Consequently, if one insists on using a minimal
set of color directions, only one color matrix $Q$, along with its
negative $-Q$, is necessary to construct all possible vortex configurations.
Without loss of generality, let
\begin{equation}
Q=\, \frac{2\pi }{3} \, \mbox{diag} \, (1,1,-2) \ .
\end{equation}
Now, consider a vortex branching. The only way to divide an incoming vortex
flux $Q$ into two outgoing fluxes is by letting the latter each be
associated with the flux $-Q$. Otherwise, evaluating Wilson loops encircling
the incoming vortex or both the outgoing vortices, respectively, would
yield different results, violating the Bianchi constraint (continuity of
flux modulo $2\pi$). However, even in the allowed branching configuration,
the incoming flux is not simply the sum of the outgoing fluxes; the flux
matrix is discontinuous. There is a sink of flux $3Q$ at the branching
(which is compatible with the Bianchi constraint); this is an Abelian
magnetic monopole\footnote{Note furthermore that, in this particular
case, branchings are the {\em only} locations where monopoles can be
present. A switch from a flux $Q$ to a flux $-Q$ along a single vortex
does not correspond to a monopole allowed by the Bianchi constraint. This
is different from the $SU(2)$ case, where a switch from
$Q_{SU(2)} = \pi \, \mbox{diag} \, (1,-1)$ to $-Q_{SU(2)} $
is possible. In the minimal description of $SU(3)$ vortices, therefore,
the vortex branching and the Abelian monopole concepts can be used
synonymously. Note that this strict identification does not persist for
$SU(4)$; in that case, monopoles can be present away from branchings even
in the minimal description.}.

As a consequence, one can also take an alternative stance. Instead of
discussing branchings for $N\geq 3$ in particular, and noting that
monopoles are present at branchings, one can take the point of view that
vortex configurations in general contain monopoles, which are sources or
sinks of up to $N$ vortex fluxes. Monopoles are connected by vortices,
and between monopoles, vortex fluxes are oriented. Note that, in this
language, there is no conceptual distinction between the cases $N=2$ and
$N\geq 3$. Both cases are described in the same language. In the former
case, only two vortices emanate from any monopole, whereas for higher $N$,
there can be two or more vortices emanating from monopoles, cf.~also
Fig.~\ref{monopfig} below. This subsumes the statement that vortices
branch for $N\geq 3$, while maintaining the same conceptual framework
for all $N$.

\subsection{Nexus construction}
Alternative ways to describe the color structure of vortex branchings
result if one allows for a non-minimal set of color flux matrices. In
particular, if one uses a sufficiently large set of different flux
matrices $Q$ along with their negatives $-Q$, one can completely
disassociate Abelian monopoles from branchings. Such a description of
branchings was introduced in \cite{cw1,cw2,cw3} along with the term ``nexus''
to refer to these objects\footnote{More precisely, these are referred to as
``quasi-Abelian'' nexi in \cite{cw1,cw2,cw3}. On the other hand, the ``fully
non-Abelian'' nexi of \cite{cw1,cw2,cw3} correspond, after a (singular) gauge
transformation, to the ``Abelian monopoles'' of the present work.}. Staying
with the $SU(3)$ example, the relevant flux matrices are
\begin{equation}
\frac{2\pi }{3} \, \mbox{diag} \, (1,1,-2) \ , \ \ \
\frac{2\pi }{3} \, \mbox{diag} \, (1,-2,1) \ , \ \ \
\frac{2\pi }{3} \, \mbox{diag} \, (-2,1,1)
\end{equation}
(along with their negatives). By letting each of the three vortices
meeting at a branching be associated with a different flux matrix,
monopoles can be completely avoided at branchings\footnote{This property
is likely to single out the quasi-Abelian nexus construction of branchings
as the one most suited in practice for the purpose of evaluating the
topological charge of general $SU(3)$ vortex world-surface configurations;
this is currently under investigation.}. Of course, the global structure
of the vortex world-surfaces will in general force monopoles to be present
elsewhere (which is consistent with the nexus description, in
contradistinction to the minimal description discussed above).

On the other hand, note that the nexus language distinguishes the $N=2$ and
the $N\geq 3$ cases on a qualitative level. The former case does not allow
for (quasi-Abelian) nexi, i.e., branchings, whereas all other cases do
\cite{cw1}.

\subsection{Dynamical issues}
\label{dyniss}
Whether one views vortex color structure in terms of Abelian monopoles
alone (as discussed at the end of section \ref{minconst}) or chooses to
differentiate between nexi (i.e., branchings) and monopoles, these
objects embody a nontrivial color structure which suggests an alternative
characterization of the Yang-Mills vortex vacuum. Instead of
viewing vortex world-surfaces as the only dynamical degrees of freedom,
which happen to force the presence of certain color structures through
their geometrical properties (nonorientability, branching), one can
alternatively view monopoles (and nexi) as bona fide degrees of freedom,
which, at least to a certain extent, determine the form of the vortex
fluxes emanating from them and connecting them. Which of these pictures
is the more appropriate one is a dynamical question; phrased more formally,
is the vortex effective action dominated by terms involving geometrical
vortex world-surface characteristics or by terms involving monopole
(and nexus) characteristics?

\begin{figure}[h]
\centerline{\epsfig{file=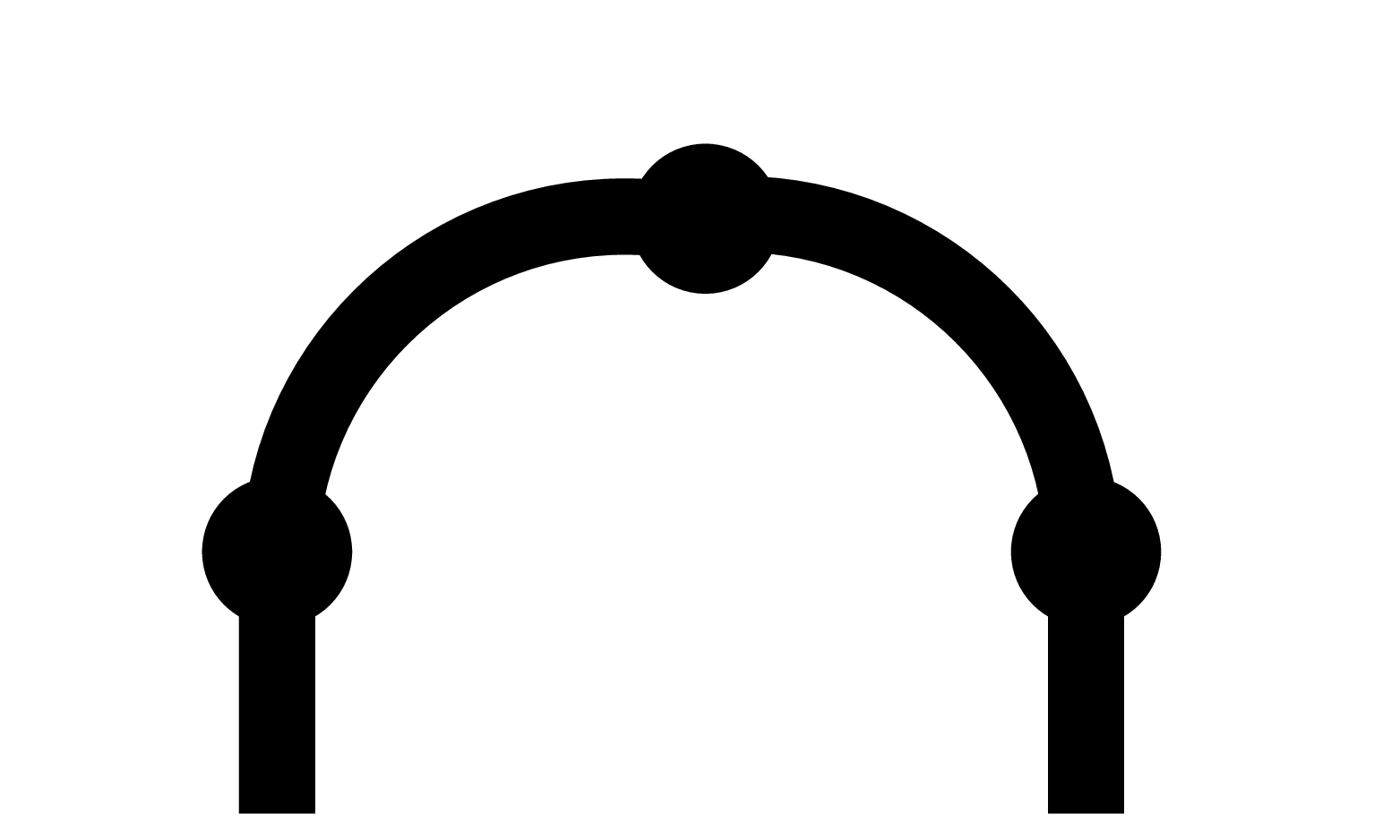,width=4cm} \hspace{2cm}
\epsfig{file=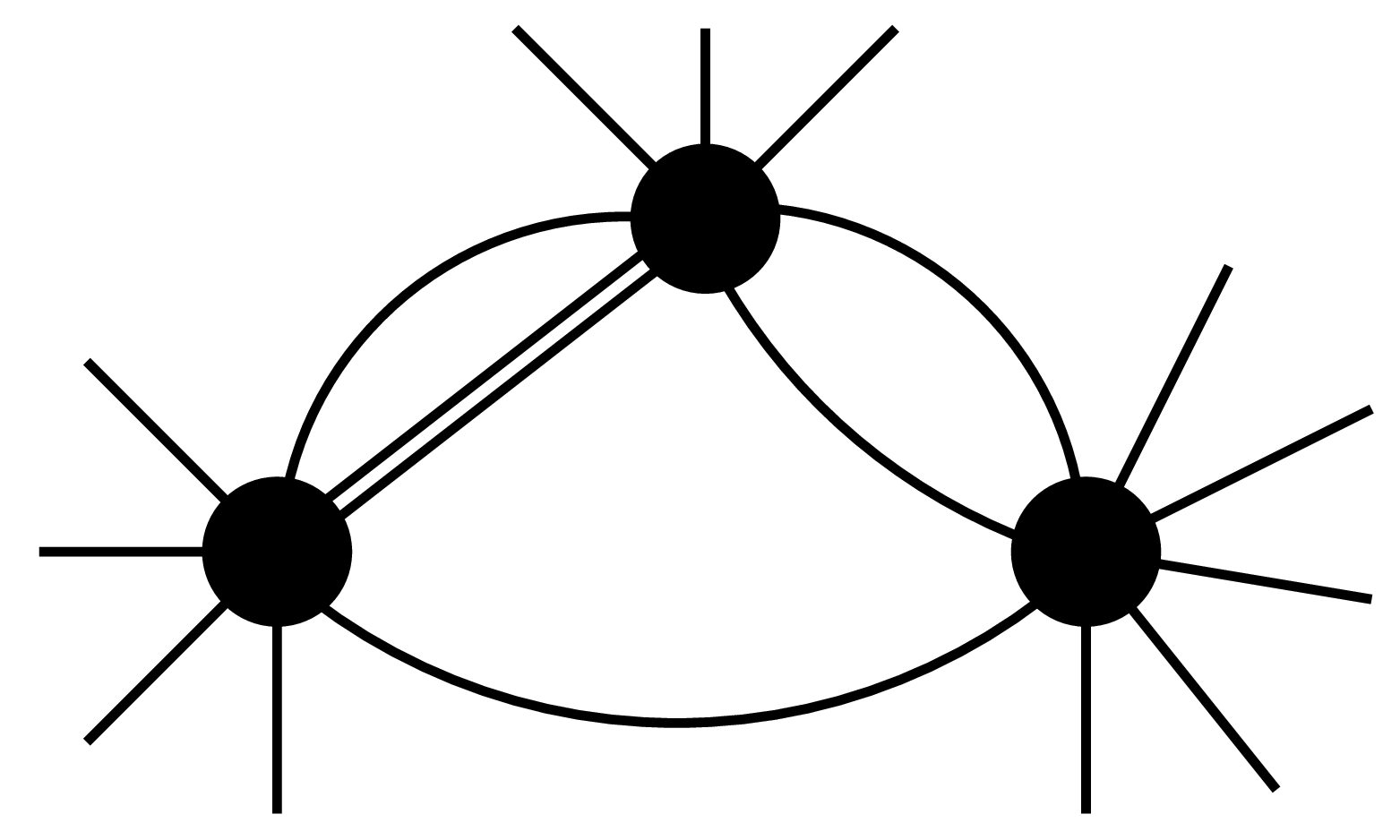,width=4cm}}
\caption{Vortex configurations contain monopoles, cf.~section \ref{minconst}.
For $N$ colors, up to $N$ vortex fluxes can emanate from any given monopole.
Examples for $N=2$ (left) and $N=8$ (right) are depicted. In order to
display a generic case, for $N=8$ also a vortex associated with center phase
$e^{iQ} = e^{i\pi /2} $ (the double line) has been included (all the
other vortices carry center phase $e^{iQ} = e^{i\pi /4} $). As $N$ rises,
center flux can be quantized in ever smaller units, while monopoles
remain sources or sinks of flux $2\pi $, or multiples thereof (in each
diagonal color component).}
\label{monopfig}
\end{figure}

In the $SU(2)$ and $SU(3)$ cases studied to date, effective vortex actions
based purely on world-surface characteristics appear entirely adequate to
reproduce the infrared sectors of the respective Yang-Mills theories on a
quantitative level. Specifically, vortices are controlled by world-surface
curvature in the $SU(2)$ and $SU(3)$ models.
However, it would be rather surprising if this picture persisted
indefinitely as the number of colors $N$ is increased; on the contrary,
a shift towards dynamics influenced by monopole (or nexus) characteristics
is likely, as has been argued previously \cite{jeffstef}. This is
expected because, as the number of colors rises, center flux can be quantized
in ever smaller units, while the monopoles always constitute sources or
sinks of flux $2\pi $, or multiples thereof (in each diagonal color
component), cf.~Fig.~\ref{monopfig}. Thus, it seems plausible that
monopoles become the dominant carriers of inertia in vortex configurations,
and that the vortex world-surfaces then adjust to the monopole positions
rather than the latter being determined by the world-surface dynamics.
This should manifest itself in an effective vortex action in which the
monopoles (or nexi) attain their own dynamical significance.

To search for indications of this shift in the dynamical characteristics,
and thus probe the limits of applicability of pure random world-surface
dynamics for center vortices, was the principal motivation for the present
$SU(4)$ work. Indeed, of the effective vortex actions investigated below, only
ones including an explicit dependence on branching properties prove to be
phenomenologically viable. Before proceeding to describe the investigation in
detail, it should be emphasized that the possibility of such a shift in the
dynamical characteristics does not imply that the vortex picture as a whole
may become inappropriate at larger $N$ and that, e.g., an Abelian monopole
Coulomb gas may become an appropriate description. On the contrary, for any
finite $N$, chromomagnetic flux must be constricted into center vortices in
order to, e.g., correctly account for the lack of confinement of adjoint
color sources \cite{jg3}. The aforementioned shift in the
characterization of the vacuum merely concerns the specific form
of the effective vortex action; vortices nonetheless continue to represent
relevant infrared degrees of freedom. All that is implied is that,
at larger $N$, the Abelian
monopoles which intrinsically reside on generic non-orientable vortex
world-surfaces (in Abelian gauges) may cease to be completely dependent
objects with no dynamical significance of their own; instead, their
characteristics may begin to play a role in determining overall vortex
configuration dynamics\footnote{Merely at infinite $N$, where the center
of the $SU(N\rightarrow \infty )$ gauge group becomes
$Z(N\rightarrow \infty ) \simeq U(1)$, a pure monopole description
may be feasible. There, center quantization of flux ceases to constitute
a constraint, since arbitrarily small units of center flux are possible.}.

\section{SU(4) vortex model}
\subsection{Degrees of freedom}
Vortex flux quantization in the $SU(4)$ vortex model is determined by
the three nontrivial center elements $i,-i$ and $-1$ of the $SU(4)$ gauge
group. The former two elements are complex conjugates of one another and
thus can be generated by the two possible orientations of one type of
vortex. Therefore, the $SU(4)$ model contains in all two physically
different types of vortices; the ones generating a phase $-1$ and the ones
generating the phases $\pm i$ depending on their orientation in space-time.

The $SU(4)$ random vortex world-surface model is constructed in complete
analogy to the $SU(2)$ and $SU(3)$ cases studied previously. The reader
is referred to \cite{m1,su3conf} for details regarding the physical
interpretation of the construction. The vortex world-surfaces are modeled
by composing them of elementary squares on a hypercubic lattice. Each
elementary square in the lattice is associated with a value
\begin{equation}
q_{\mu \nu } (x) \in \{ -1,0,1,2\} \ ,
\ \ \ \ \ \ \ \ q_{\nu \mu } (x) = -q_{\mu \nu } (x)
\end{equation}
where the square extends from the point $x$ into the positive $\mu $ and
$\nu $ directions. The right-hand relation is simply a reminder of the
behavior of flux under space-time inversions, i.e., when the orientation of
the vortex surface is reversed. In practice, recording only
$q_{\mu \nu } (x)$ for $\mu < \nu $ is sufficient. The value
$q_{\mu \nu } (x) =0$ corresponds to no vortex flux being present on the
square in question; nonzero values of $q_{\mu \nu } (x)$ could, e.g., label
the first element of the flux matrix $Q_{\mu \nu } (x)$ associated with the
elementary square in a minimal Abelian construction,
\begin{equation}
Q_{\mu \nu } (x) \in \left\{
\frac{\pi }{2} \, \mbox{diag} \, (-1,-1,-1,3) \ , \
\frac{\pi }{2} \, \mbox{diag} \, (1,1,1,-3) \ , \
\frac{\pi }{2} \, \mbox{diag} \, (2,2,-2,-2) \right\}
\label{fluma}
\end{equation}
(since, in the following, only Wilson loops and action densities will
be studied, it is not necessary to distinguish between the last flux
matrix in (\ref{fluma}) and the one corresponding to inverse orientation;
for a study, e.g., of topological properties, this would be necessary).

An important point to be noted is that the lattice spacing in
this approach is a fixed physical quantity implementing the notion that
vortices possess a finite transverse thickness and must be a minimal
distance apart to be distinguished from one another\footnote{In the
$SU(2)$ and $SU(3)$ models, the lattice spacing turns out to be $0.39 $ fm
if one fixes the scale by setting the zero-temperature string tension to
$\sigma (T=0) = (440\, \mbox{MeV} )^2 $.}. This is discussed in detail in
\cite{m1,su3conf}. While in the $SU(2)$ and $SU(3)$ models, there is
only one type of center vortex (up to orientation), in the present $SU(4)$
case, there are two physically distinct types of vortices. In general,
these can have different thicknesses. However, there is no straightforward
way of implementing variable thicknesses in the present hypercubic lattice
formulation of the model; all vortices will be described using a single
lattice. This model restriction should be kept in mind.

Ensembles of random vortex world-surface configurations on lattices are
generated using Monte Carlo methods. The elementary update of a given
vortex surface configuration can be effected such that the Bianchi
constraint is respected at every step. This is achieved by updating
all six faces of an elementary three-dimensional cube in the lattice
at once, in a way which corresponds to adding a vortex flux of
the shape of the cube surface to the flux previously present. Formally,
if the elementary cube in question extends from the lattice site $x$ into
the positive $\mu $, $\nu $ and $\lambda $ directions, then update
simultaneously
\begin{equation}
\begin{array}{ll}
q_{\mu\nu}(x) \rightarrow (q_{\mu\nu}(x) + w)
\, \mbox{mod} \, 4\, , \ \ \ \ &
q_{\mu\nu}(x+e_\lambda) \rightarrow (q_{\mu\nu}(x + e_\lambda) - w)
\, \mbox{mod} \, 4 \\
q_{\nu \lambda}(x) \rightarrow (q_{\nu \lambda}(x) + w)
\, \mbox{mod} \, 4\, , \ \ \ \ &
q_{\nu \lambda}(x+e_\mu) \rightarrow (q_{\nu \lambda}(x+e_\mu) - w)
\, \mbox{mod} \, 4 \\
q_{\lambda\mu}(x) \rightarrow (q_{\lambda\mu}(x) + w)
\, \mbox{mod} \, 4\, , \ \ \ \ &
q_{\lambda\mu}(x+e_\nu) \rightarrow (q_{\lambda\mu}(x+e_\nu) - w)
\, \mbox{mod} \, 4
\end{array}
\end{equation}
where the value $w\in \{-1,1,2 \} $ characterizing the superimposed flux
in practice is chosen at random with equal probability, and the modulo
operation is to be carried out such that the result again satisfies
$q\in \{-1,0,1,2 \} $. Since the update effects a linear superposition
of two fluxes which satisfy the Bianchi constraint, the updated
configuration again satisfies that constraint (the modulo operation
merely generates shifts by $2\pi $ in the flux matrix, which are allowed
by the Bianchi constraint; in this way, monopoles appear in the vortex
configuration as the updates proceed).

\subsection{Action}
Finally, the action weighting vortex world-surface configurations must
be specified. The most general form of the action used is the sum
\begin{equation}
S[q] = S_{\rm curv}[q] + S_{\rm area}[q] + S_{\rm branch}[q]
\end{equation}
with the individual terms specified as follows.

The immediate generalization of the dynamics studied in the
$SU(2)$ and $SU(3)$ models is a world-surface curvature action,
adapted to the present case, in which there are two physically distinct
types of vortices,
\begin{eqnarray}
S_{\rm curv}[q] \! \! \! \! &=& \! \! \! \!
\sum_x\sum_\mu \Bigg[ \sum_{\nu < \lambda \atop \nu \neq \mu,
\lambda\neq \mu} \Big( C(| q_{\mu\nu}(x) \, q_{\mu\lambda}(x) |)
 + C(| q_{\mu\nu}(x) \, q_{\mu\lambda}(x-e_\lambda) |) \label{curvature} \\
& & \ \ \ \ \ \ \ \ \ \ \ \ \ \ \ \ \ \
+ C(| q_{\mu\nu}(x-e_\nu) \, q_{\mu\lambda}(x) |)
 + C(| q_{\mu\nu}(x-e_\nu) \, q_{\mu\lambda}(x-e_\lambda) |)
\Big)\Bigg] \nonumber
\end{eqnarray}
with
\begin{equation}
C(0)=0 \ \ \ \ C(1)=c_{11} \ \ \ \ C(2)=c_{12} \ \ \ \ C(4)=c_{22} \ .
\end{equation}
Note that this action term (as well as all others below) treats $q=\pm 1$
fluxes symmetrically, as it should,
since these two cases correspond to the two possible orientations of the
same physical vortex type. As is clear from the expression given, for
each link extending from the site $x$ into the positive $\mu $ direction,
all pairs of elementary squares attached to that link are examined,
excluding those in which the two squares lie in the same plane.
Depending on the flux associated with the squares, an action increment is
incurred. Thus, world-surfaces are penalized for going around a corner,
i.e., for curvature. In the following, the couplings $c_{ij}$
will be generated from two independent coefficients $c_1 , c_2 $
(corresponding to the two physically distinct types of vortices) as
\begin{equation}
c_{ij} = c_i c_j \ .
\label{curfact}
\end{equation}
Note that giving $c_{12} $ special treatment, i.e., generalizing to
$c_{12} \neq c_1 c_2 $, to an extent would amount to a special treatment of
branchings (along with related features such as different vortex types
intersecting along a whole line), since branchings are locations where
a $q=2$ vortex splits into two $q=\pm 1$ vortices. However, on the other
hand, an additional action term weighting branchings {\em explicitly} will
be introduced separately below. For this reason, the option
$c_{12} \neq c_1 c_2 $ was not pursued further in this investigation
(even though, strictly speaking, it is physically distinct from the
aforementioned branching action). Instead, the explicit branching action
term introduced below will be studied extensively.

A further type of action investigated in the $SU(2)$ and $SU(3)$ models is
an action simply weighting world-surface area,
\begin{equation}
S_{\rm area}[q] = \sum_x \sum_{\mu < \nu } A(|q_{\mu \nu } (x)|) \ .
\label{aract}
\end{equation}
In the $SU(2)$ and $SU(3)$ models, it was found that this action term could
be traded off against the curvature term, and it was therefore ultimately
discarded. Essentially, in the random surface ensemble, total surface area
is strongly correlated with total curvature content, such that, for all
practical purposes, the area term can be replaced by a strengthening of
the curvature term for a wide range of parameters \cite{m1,su3conf}.
In the present work, the effects of this term are briefly explored,
cf.~section~\ref{areact} below; inclusion of this term however does not
appear to increase the phenomenological flexibility of the model.

Finally, a new term introduced in the present investigation is an explicit
weighting of branchings,
\begin{equation}
S_{\rm branch}[q] = \sum_x \sum_{\mu } B \left( \sum_{\nu \neq \mu }
E(q_{\mu\nu}(x)) + E(q_{\mu\nu}(x-e_\nu)) \right)
\label{sbranch}
\end{equation}
where
\begin{eqnarray}
E(0)=0 \ , & & E=1 \ \ \mbox{else} \\
B(3)=B(5)=-b \ , & & B=0 \ \ \mbox{else}
\end{eqnarray}
This action term precisely encourages vortex branchings: For each link
extending from the site $x$ into the positive $\mu $ direction, the number
of attached elementary squares occupied by vortex flux is counted and, if
that number equals 3 or 5 (which happens precisely when branching takes
place), the action is decremented by the branching coefficient $b$.
Further below, it will be shown why the introduction of this type of term
is necessary to arrive at a phenomenologically viable model, corroborating
the general arguments of section \ref{motiv}. In view of the different
vortex formulations discussed in section \ref{motiv}, one can interpret
$S_{\rm branch} $ as favoring the presence of nexi, or alternatively as
favoring the presence of (certain types of) Abelian monopoles. Whichever
picture one may choose, the introduction of this term represents a
departure from pure vortex world-surface dynamics, according additional
color structures present on the vortex world-surfaces their own dynamical
significance.

\section{SU(4) Yang-Mills lattice data}
\label{datasec}
The $SU(4)$ lattice data which will be matched in the random vortex
world-surface model are drawn from \cite{luctep03,luctep04}. After the
bulk of the numerical work in the present investigation was completed,
updated lattice data became available \cite{luctepnew}. The author did
not attempt to adjust the vortex model to the revised data, since the
quantitative adjustments this would entail do not alter the conclusions
drawn.

On the other hand, \cite{luctepnew} also contains lattice data on the
spatial string tension, for which no measurements were available previously.
These data will be compared to corresponding predictions within the vortex
model constructed here in section \ref{bransec}.

Specifically, the $SU(4)$ Yang-Mills characteristics which will be matched
in the vortex model are the following:

\subsection{String tension and deconfinement temperature}
The ratio of the deconfinement temperature $T_c $ to the zero-temperature
quark string tension $\sigma_{1} $ is taken to be \cite{luctep03}
\begin{equation}
\frac{T_c }{\sqrt{\sigma_{1} } } = 0.62 \ .
\label{tcds}
\end{equation}
The quantity $T_c /\sqrt{\sigma } $ was also used in the $SU(2)$ and $SU(3)$
models to fix the single curvature coefficient $c$; since that curvature
coefficient represented the only (physically significant) adjustable
parameter, no further lattice data were needed to fully define those models.

\subsection{String tension ratios}
In correspondence to the two different types of center vortices present in
the $SU(4)$ theory, there are also two distinct string tensions, the
quark string tension $\sigma_{1} $ and the diquark string tension
$\sigma_{2} $. Data on both is needed to fix the two curvature coefficients
$c_1 , c_2 $ of the $SU(4)$ vortex model. Thus, as one generalizes the
vortex model to a higher number of colors, it does not generally predict
ratios between different string tensions, unless one makes further prior
assumptions about relationships between the properties of different types
of vortices; for such considerations at large $N$, cf.~\cite{jeffstef}.
Rather, the vortex model always provides sufficient freedom to match
all relevant string tension ratios, since there are correspondingly many
physically distinct types of vortices with their separate dynamical
characteristics such as curvature coefficients, etc.

The ratio of the diquark string tension to the quark string tension at
zero temperature is taken to be \cite{luctep04}
\begin{equation}
\frac{\sigma_{2} }{\sigma_{1} } = 1.36 \ .
\label{ratin}
\end{equation}

\subsection{First order character of the deconfinement transition}
The first-order character of the deconfinement phase transition becomes
more pronounced as one raises the number of colors. In units of $T_c^4 $,
the latent heat of $SU(4)$ Yang-Mills theory is larger than the one of
$SU(3)$ Yang-Mills theory by a factor of 2.25 \cite{luctep03}. As will
be described below, the $SU(4)$ vortex model with a pure curvature
action (\ref{curvature}) does not reproduce this property, even on
a qualitative level. This is the reason for the introduction of the new
branching action (\ref{sbranch}); a more detailed motivation for that
particular choice follows further below. Introducing this additional
action term introduces the new branching coefficient $b$; to fix this
coefficient, an additional piece of lattice data is needed. The
aforementioned relation between latent heats in the $SU(3)$ and
$SU(4)$ cases can be restated as
\begin{equation}
\left. \Delta s \frac{T_c^4 }{\sigma_{1}^{4} } \right|_{SU(4)} =
2\cdot \left. \Delta s \frac{T_c^4 }{\sigma^{4} } \right|_{SU(3)}
\label{input2}
\end{equation}
using (\ref{tcds}) as well as the relation $T_c /\sqrt{\sigma } =0.63$ for
$SU(3)$ used in \cite{su3conf}, where $\Delta s $ denotes the difference
in action density (per space-time volume) between the two coexisting phases
at the deconfinement temperature. This modified choice of units was adopted
because it has more moderate scaling properties as one investigates the
vortex model on lattices of different (Euclidean) temporal extent $N_t $,
cf.~below.

In order to be able to model the property (\ref{input2}), it is necessary
to briefly revisit the $SU(3)$ vortex model \cite{su3conf}. The case of
lattices with an extension of $N_t =2$ lattice spacings in the Euclidean
time direction is discussed in detail in \cite{su3conf}; the deconfinement
transition is found at the value $c=0.2359$ for the curvature coefficient.
Reading off the discontinuity in the action density at the phase transition
from the corresponding action density distribution\footnote{Note that
the distributions in \cite{su3conf} are plotted as a function of the
action per link divided by the curvature coefficient $c$; therefore, to
obtain $\Delta s $, one needs to multiply by $4c/a^4$, where $a$ denotes
the lattice spacing. Note also that the rough measure for the action density
discontinuity used here and throughout this work is simply the distance
between the two maxima in the action density distribution plot (instead
of, e.g., modeling the latter as a superposition of two distributions).
Since only ratios between discontinuities are of interest here, the effects
of this slight underestimate cancel to a sufficient degree for the present
purposes.}, one has (with $a$ denoting the lattice spacing)
\begin{equation}
\Delta s = 0.45 \, T_c^4 \ \ \ \ \ \ \ \ \
(T_c a = 1/2, \ c=0.2359, \ \sigma a^2 = 0.695)
\label{ds2}
\end{equation}
where the value for the zero-temperature string tension $\sigma $ measured
separately for $c=0.2359$ has been included for later reference.
The physical point of the $SU(3)$ model is not at $c=0.2359$, but nearby
at $c=0.21$. This can be corrected for by also considering $N_t =1$ and
interpolating in $c$. For $N_t =1$, the deconfinement transition is found
for $c=0.0872$. Producing an action density distribution analogous to the
one discussed above for $N_t =2$ permits one to read off
\begin{equation}
\Delta s = 0.022 \, T_c^4 \ \ \ \ \ \ \ \ \
(T_c a = 1, \ c=0.0872, \ \sigma a^2 = 1.355) \ .
\label{ds1}
\end{equation}
The zero-temperature string tension $\sigma $ was again measured separately
for this value of $c$. From (\ref{ds2}) and (\ref{ds1}), the strong
variation of $\Delta s$ as a function of $c$ in the units given is evident.
As a consequence, these units are not well suited for interpolation in $c$,
and it is indeed advantageous to convert to
\begin{equation}
\left. \Delta s \frac{T_c^4 }{\sigma^{4} } \right|_{c=0.2359} = 
0.0075 \ , \ \ \ \ \ \ \ \ \ \ \
\left. \Delta s \frac{T_c^4 }{\sigma^{4} } \right|_{c=0.0872} =
0.0065 \ .
\end{equation}
Finally, interpolating linearly to the physical point $c=0.21$, one has
\begin{equation}
\left. \Delta s \frac{T_c^4 }{\sigma^{4} } \right|_{SU(3)} = 0.0074
\end{equation}
and, inserting into (\ref{input2}), the condition to be satisfied by the
$SU(4)$ model is therefore
\begin{equation}
\left. \Delta s \frac{T_c^4 }{\sigma_{1}^{4} } \right|_{SU(4)} = 0.015 \ .
\label{latin}
\end{equation}

\begin{figure}[ht]
\centerline{\epsfig{file=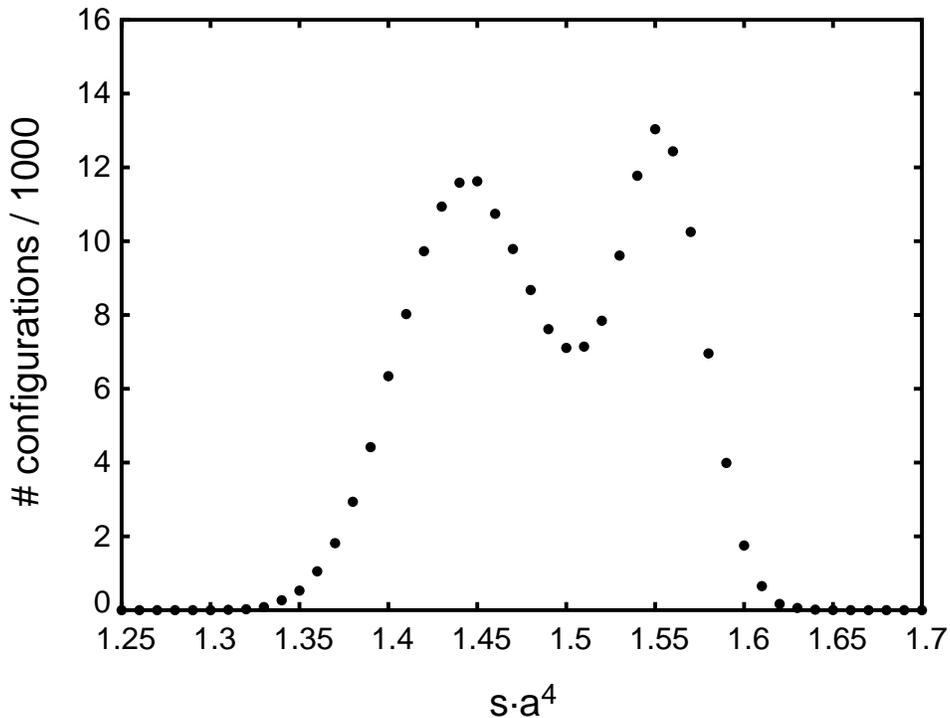,width=13cm} }
\caption{Distribution of the action density $s$ (per space-time volume)
at the deconfining phase transition, for a pure vortex world-surface
curvature action with $c_1 =c_2 =0.4722$. The measurement was taken on a
$14^3 \times 2$ lattice; $a$ denotes the lattice spacing.}
\label{c2ec1}
\end{figure}

\section{Exploration of coupling parameter space}
\subsection{Pure world-surface curvature action}
\label{simpmod}
As a starting point, consider the $SU(4)$ random vortex world-surface model
with a pure curvature action (\ref{curvature}),(\ref{curfact}), restricted
to equal curvature coefficients, $c_1 =c_2 $. This is not yet very realistic
as far as the condition (\ref{ratin}) is concerned; for $c_1 =c_2 $, one
obtains $\sigma_{1} =\sigma_{2} $, i.e., $\sigma_{2} /\sigma_{1} =1$. On
the other hand, one does observe a deconfinement transition which is quite
strongly first order, cf.~Fig.~\ref{c2ec1} \, for $N_t =2$, taken at
$c_1 =c_2 =0.4722$. Similar observations can be made at other $N_t $,
with corresponding $c_1 =c_2 $; the case $N_t =2$ is the one which comes
closest to satisfying the remaining condition (\ref{tcds}). Note that the
physical set of parameters in the random vortex world-surface model is
generally such that the inverse deconfinement temperature is not an integer
multiple of the lattice spacing; properties of the deconfinement transition
at the physical point then cannot be measured directly, but are inferred by
interpolation of data obtained at different $N_t $, cf.~the treatment of the
$SU(3)$ case above and cf.~also \cite{m1,su3conf}.

\begin{figure}[ht]
\centerline{\epsfig{file=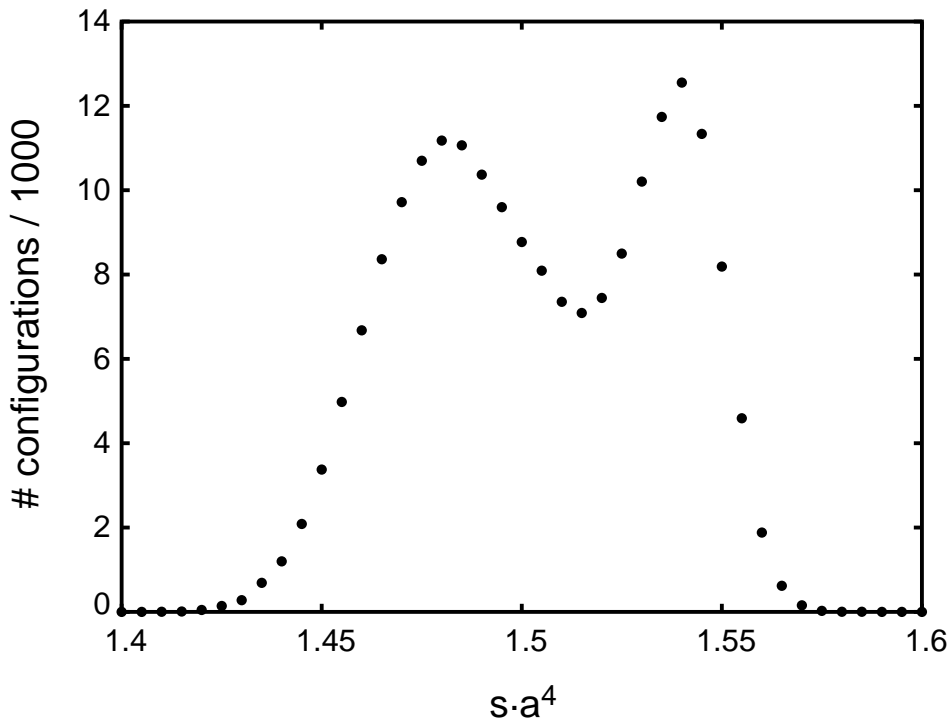,width=13cm} }
\caption{Distribution of the action density $s$ at the deconfining phase
transition, for a pure vortex world-surface curvature action with
$c_1 =0.4537$ and $c_2 =0.5037$. The measurement was taken on a
$26^3 \times 2$ lattice.}
\label{c2gc1}
\end{figure}

As a next step, in order to approach the condition (\ref{ratin}), it is
necessary to increase $c_2 $ over $c_1 $. For example, the set of curvature
coefficients $c_1 =0.4537$, $c_2 =0.5037$ yields
$\sigma_{2} /\sigma_{1} =1.17$ while simultaneously realizing the
deconfinement transition on a $N_t =2$ lattice, cf.~Fig.~\ref{c2gc1}.
Examining the trends, one notices an unexpected behavior: As one attempts
to modify the parameters to approach the condition (\ref{ratin}), the
first-order character of the phase transition is weakened. In fact, when
(\ref{ratin}) is finally satisfied in this way, the action discontinuity
at the phase transition has either become undetectably small or has
disappeared entirely, rendering the deconfinement transition second order.
To be specific, the author used string tension measurements on $N_t =2$
lattices in addition to the zero-temperature measurements to locate the
set of parameters $c_1 =0.428$, $c_2 =0.540$ which yields (\ref{ratin})
as well as realizing the deconfinement transition at $N_t =2$; then, a
careful scan of $c_1 $ in steps of $0.0001$ around $c_1 =0.428$ was
performed (with $c_2 =c_1 +0.112$ in each case), generating the action
density distributions on large ($30^3 \times 2$) lattices. No evidence of
first order behavior was found.

This qualitative behavior can be confirmed for other $N_t $. For $N_t =1$,
the deconfinement transition occurs at $c_1 =0.253$, $c_2 =0.353$ (with
(\ref{ratin}) also satisfied). Likewise, for $N_t =3$, the transition
occurs at $c_1 =0.538$, $c_2 =0.571$. In both cases, a careful scan in
$c_1 $, as above, with $c_2 -c_1 $ fixed, reveals no first-order behavior.
By extension, also the phase transition at the interpolated set of
parameters realizing the remaining condition (\ref{tcds}) is not
discernibly first order.

Thus, it must be concluded that the conditions (\ref{ratin}) and
(\ref{latin}) are incompatible as long as one restricts oneself to a 
pure world-surface curvature action. One can realize either the one
or the other condition, but one cannot model both simultaneously.

Before seeking to remedy this by generalizing the action, it is nevertheless
useful to record here, for later reference, the simplified model which
results if one disregards the first-order character of the deconfinement
phase transition, i.e., if one disregards (\ref{latin}), and uses a pure
world-surface curvature action. The two curvature coefficients $c_1 $ and
$c_2 $ are then fixed using only the two conditions (\ref{ratin}) and
(\ref{tcds}) as follows. Table~\ref{simptab} summarizes again the pairs of
curvature coefficients at which the deconfinement transition occurs for
$N_t =1,2,3$ while (\ref{ratin}) is simultaneously satisfied.

\begin{table}[h]
\begin{center}
\begin{tabular}{|c||c|c||c|}
\hline
& $c_1 $ & $c_2 $ & $T_c /\sqrt{\sigma_{1} } $ \\
\hline \hline
$N_t =1$ & 0.253 & 0.353 & 0.85 \\
\hline
$N_t =2$ & 0.428 & 0.540 & 0.60 \\
\hline
$N_t =3$ & 0.538 & 0.571 & 0.50 \\
\hline
\end{tabular}
\end{center}
\caption{Sets of coupling constants realizing the deconfinement temperature
as well as satisfying (\ref{ratin}). Condition (\ref{tcds}) is then satisfied
by interpolating the parameters using the data in the final column.}
\label{simptab}
\end{table}

The table furthermore records the ratio $T_c /\sqrt{\sigma_{1} } $
corresponding to each parameter set. To obtain
the physical point, one constructs the quadratic interpolations of the
curvature coefficients $c_1 $ and $c_2 $ as functions of
$T_c /\sqrt{\sigma_{1} } $. Setting $T_c /\sqrt{\sigma_{1} } $ equal to
its physical value, cf.~(\ref{tcds}), yields the ``physical'' set of
curvature coefficients
\begin{equation}
c_1 = 0.41 \ \ \ \ \ \ \ \ \ \ \ \ \ \
c_2 = 0.53 \ .
\label{simppar}
\end{equation}
Since the ratio $\sigma_{2} /\sigma_{1} $ satisfies (\ref{ratin}) for all
parameter sets used in the interpolation, it is expected to satisfy that
condition also at the physical point (\ref{simppar}). On the other hand,
this quantity can also be evaluated directly at the physical point,
providing a cross-check of the interpolation procedure. Measuring
$\sigma_{2} /\sigma_{1} $ directly for the parameters (\ref{simppar})
yields the value $\sigma_{2} /\sigma_{1} =1.37$, deviating only slightly
from (\ref{ratin}). This is adequate for the present simplified model,
which is only constructed here for comparison purposes. Indeed,
in the comprehensive model discussed further below, the analogous
cross-check of the interpolation procedure yields complete agreement
with (\ref{ratin}) at the physical point.

Having determined the ``most physical'' set of curvature coefficients
(\ref{simppar}) within this restricted model framework, one can predict the
behavior of the string tensions at finite temperatures. The most significant
quantitative prediction which is accessible in this way is the behavior
of the {\em spatial} quark and diquark string tensions, $\sigma_{1}^{S} $
and $\sigma_{2}^{S} $, in the deconfined phase. For the parameter set
(\ref{simppar}), the deconfined phase is realized for $N_t =1$, which
corresponds to the temperature\footnote{Note that string tensions at other
temperatures above $T_c $ are not directly accessible at (\ref{simppar}),
but can be obtained using interpolations of measurements at other parameter
sets to the physical point (\ref{simppar}), cf.~\cite{m1}. This was not
pursued further here.} $T=1.9\, T_c $, as one can infer
by measuring the zero-temperature quark string tension $\sigma_{1} $ in
lattice units and using (\ref{tcds}). At this temperature, one obtains
\begin{equation}
\sigma_{1}^{S} (T=1.9\, T_c ) / \sigma_{1} (T=0) = 1.61 \ \ \ \ \ \ \ \ \ 
\sigma_{2}^{S} (T=1.9\, T_c ) / \sigma_{2} (T=0) = 1.46 \ .
\end{equation}
Thus, the spatial diquark string tension rises less strongly in the
deconfined phase than the spatial quark string tension; the ratio
between the two at $T=1.9\, T_c $ is reduced to
\begin{equation}
\sigma_{2}^{S} (T=1.9\, T_c ) / \sigma_{1}^{S} (T=1.9\, T_c ) = 1.25
\label{simprat}
\end{equation}
compared to the zero-temperature value $\sigma_{2} /\sigma_{1} =1.36$.

\subsection{Effect of world-surface area term}
\label{areact}
As a first attempt to generalize the action in order to obtain a more
realistic phenomenology, the effect of the world-surface area term
(\ref{aract}) was briefly explored. This type of term was already
investigated within the $SU(2)$ and $SU(3)$ models \cite{m1,su3conf}
and ultimately discarded, since it did not enhance the phenomenological
flexibility of those models. In the $SU(4)$ model, it can be used to
differentiate between the two types of vortices without abandoning
$c_1 =c_2 $. Thus, the condition $c_1 =c_2 $ was maintained, and
instead (\ref{aract}) with $A(0)=A(1)=0$ but nonvanishing $A(2)$
was introduced in the hope of maintaining a discernible first-order
deconfinement transition while approaching the condition (\ref{ratin}).

\begin{figure}[ht]
\centerline{\epsfig{file=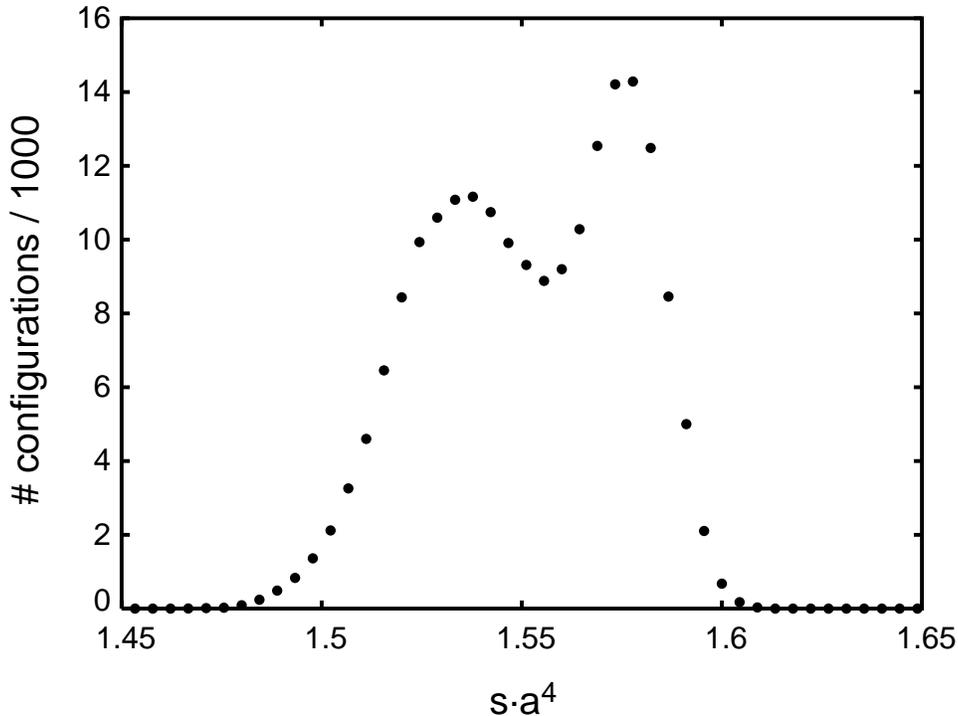,width=13cm} }
\caption{Distribution of the action density $s$ at the deconfining phase
transition, for an action containing a vortex world-surface curvature term
with $c_1 =c_2 =0.460$ and a world-surface area term with $A(2)=0.130$. The
measurement was taken on a $30^3 \times 2$ lattice.}
\label{a2ne0}
\end{figure}

As an example, using $A(2)=0.130$, the set of curvature coefficients
$c_1 =c_2 =0.460$ yields $\sigma_{2} /\sigma_{1} =1.17$ while
simultaneously realizing the deconfinement transition on a $N_t =2$
lattice, cf.~Fig.~\ref{a2ne0}. The behavior is similar to the one observed
above in the case of a pure curvature action; approaching condition
(\ref{ratin}) weakens the action density discontinuity at the phase
transition. In fact, in the present case, the discontinuity is even more
strongly suppressed than in the analogous pure curvature case at the same
ratio $\sigma_{2} /\sigma_{1} =1.17$. Thus, introducing the world-surface
area term exacerbates the problem rather than alleviating it.

Increasing $A(2)$ further, when (\ref{ratin}) is fully satisfied, the
discontinuity indeed has disappeared again. When $A(2)=0.300$, the
deconfinement transition occurs at $c_1 =c_2 =0.439$, and (\ref{ratin})
simultaneously holds. The author again carried out a careful scan of
$c_1 =c_2 $ in steps of $0.0001$ around this point and found no evidence
of first order behavior.

\subsection{Branching term}
\label{bransec}
As has become clear from the preceding sections, a vortex model action
based purely on the world-surface characteristics curvature and
area is no longer phenomenologically viable in the $SU(4)$ case. This
provides the motivation for introducing the new branching term
(\ref{sbranch}) into the vortex model action. The reason for this
particular choice lies in the following observation. Both the $SU(2)$
and the $SU(3)$ vortex models \cite{m1,su3conf} are governed by
the same type of action; nevertheless, the $SU(3)$ model exhibits a
clear first-order deconfinement transition, whereas the $SU(2)$ model
does not. The difference between the two models lies in the class of
allowed vortex topologies; only the $SU(3)$ model provides for vortex
branching. In view of this, it seems plausible to assume that facilitating
vortex branching is conducive to a first-order deconfinement phase transition.
Therefore, introducing a new action term encouraging vortex branching,
such as (\ref{sbranch}), into the $SU(4)$ vortex model is expected to
provide a viable mechanism for restoring phenomenologically correct
behavior at the phase transition. This expectation is confirmed by the
measurements discussed in the following.

\begin{figure}[ht]
\centerline{\epsfig{file=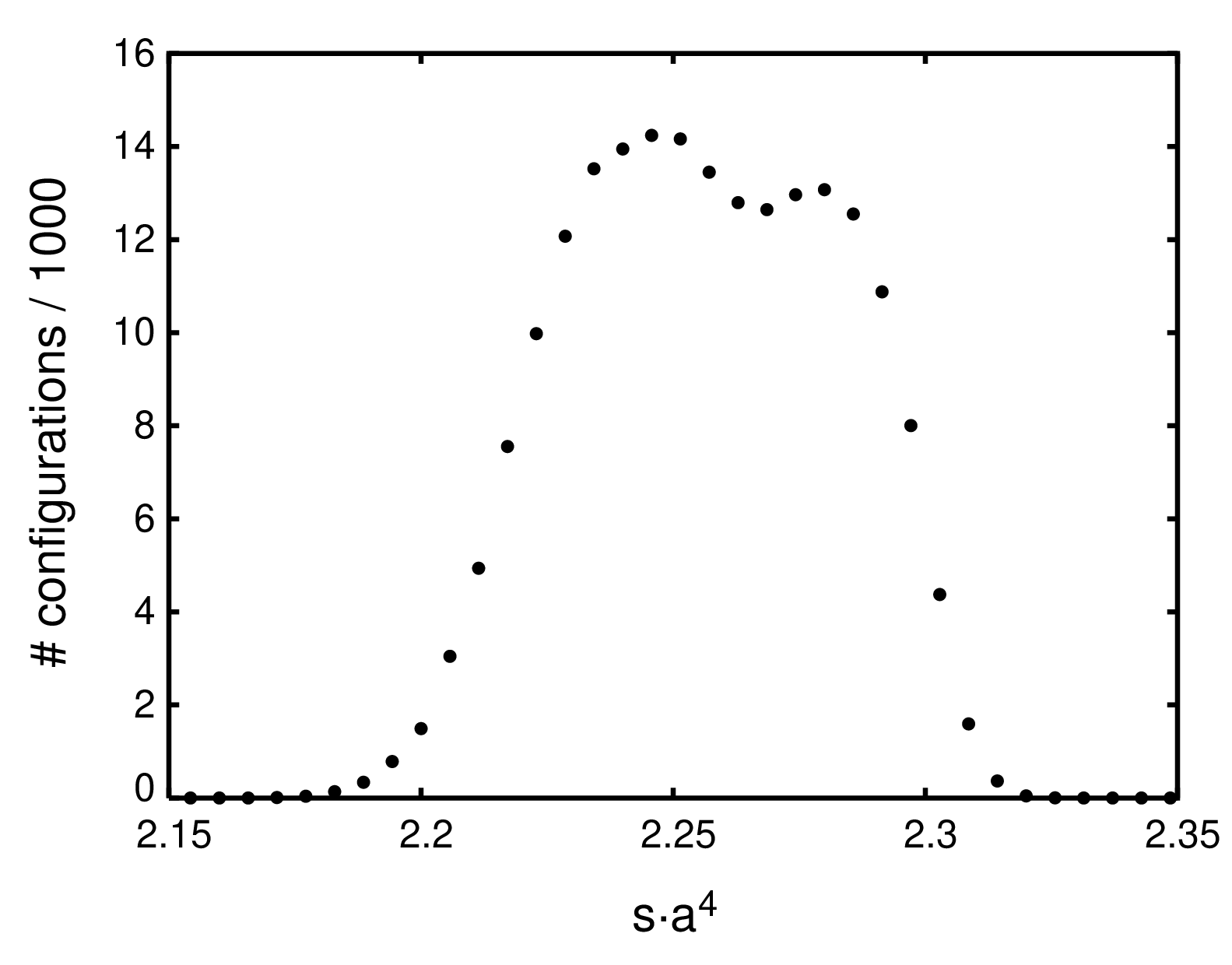,width=13cm} }
\caption{Distribution of the action density $s$ at the deconfining phase
transition, for an action containing a vortex world-surface curvature term
with $c_1 =0.2785$, $c_2 =0.4005$ and a branching term with $b=0.1403$. The
measurement was taken on a $30^3 \times 1$ lattice. Taking the distance
between the two maxima as a rough measure of the action density discontinuity
$\Delta s$ and supplementing this with a measurement of the zero-temperature
quark string tension $\sigma_{1} $ in lattice units at the same set of
coupling parameters, the condition (\ref{latin}) is satisfied by this
first-order transition.}
\label{bnt1}
\end{figure}

\begin{figure}[ht]
\centerline{\epsfig{file=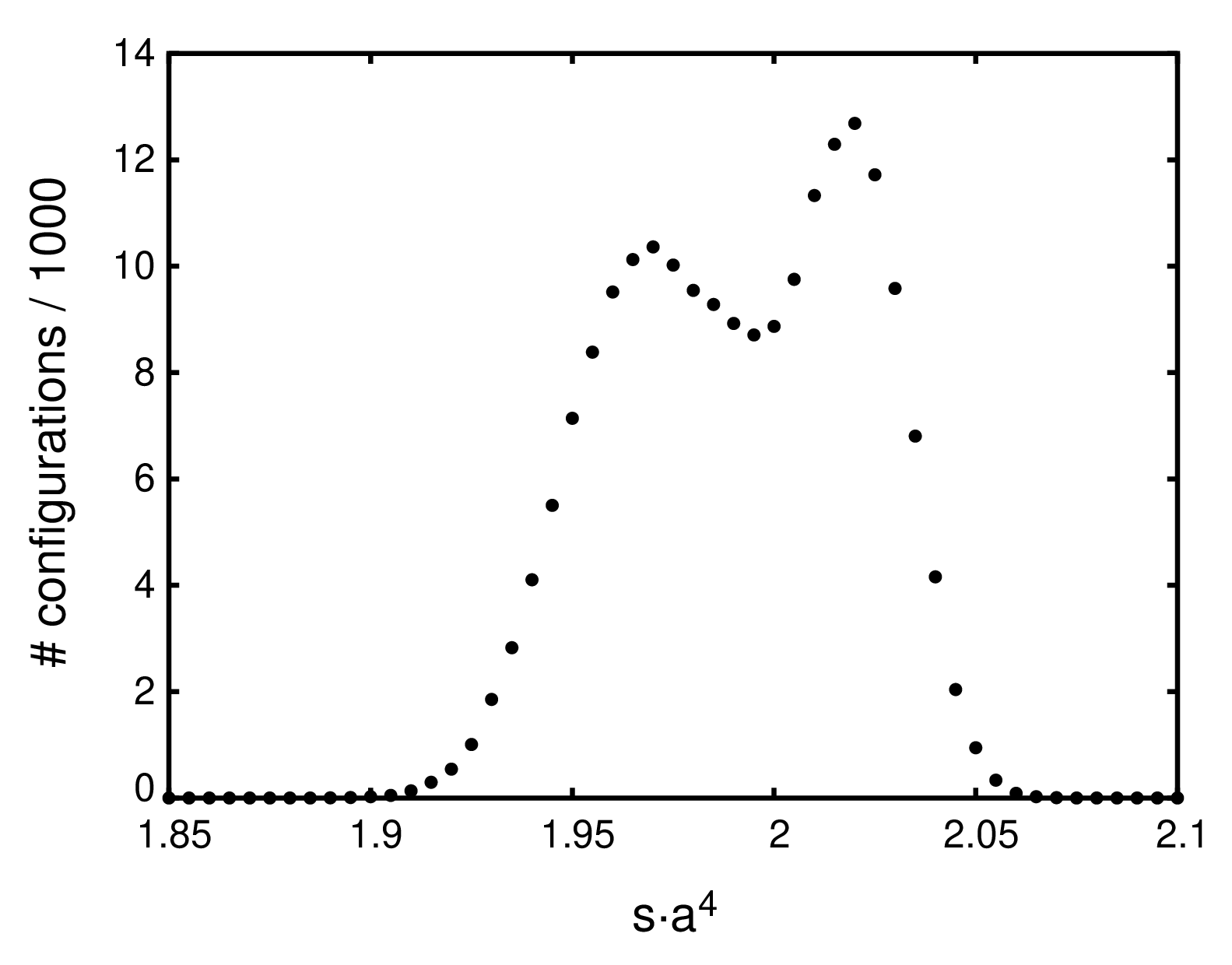,width=13cm} }
\caption{Distribution of the action density $s$ at the deconfining phase
transition, for an action containing a vortex world-surface curvature term
with $c_1 =0.4558$, $c_2 =0.7983$ and a branching term with $b=0.6950$. The
measurement was taken on a $18^3 \times 2$ lattice. Taking the distance
between the two maxima as a rough measure of the action density discontinuity
$\Delta s$ and supplementing this with a measurement of the zero-temperature
quark string tension $\sigma_{1} $ in lattice units at the same set of
coupling parameters, the condition (\ref{latin}) is satisfied by this
first-order transition.}
\label{bnt2}
\end{figure}

\begin{figure}[ht]
\centerline{\epsfig{file=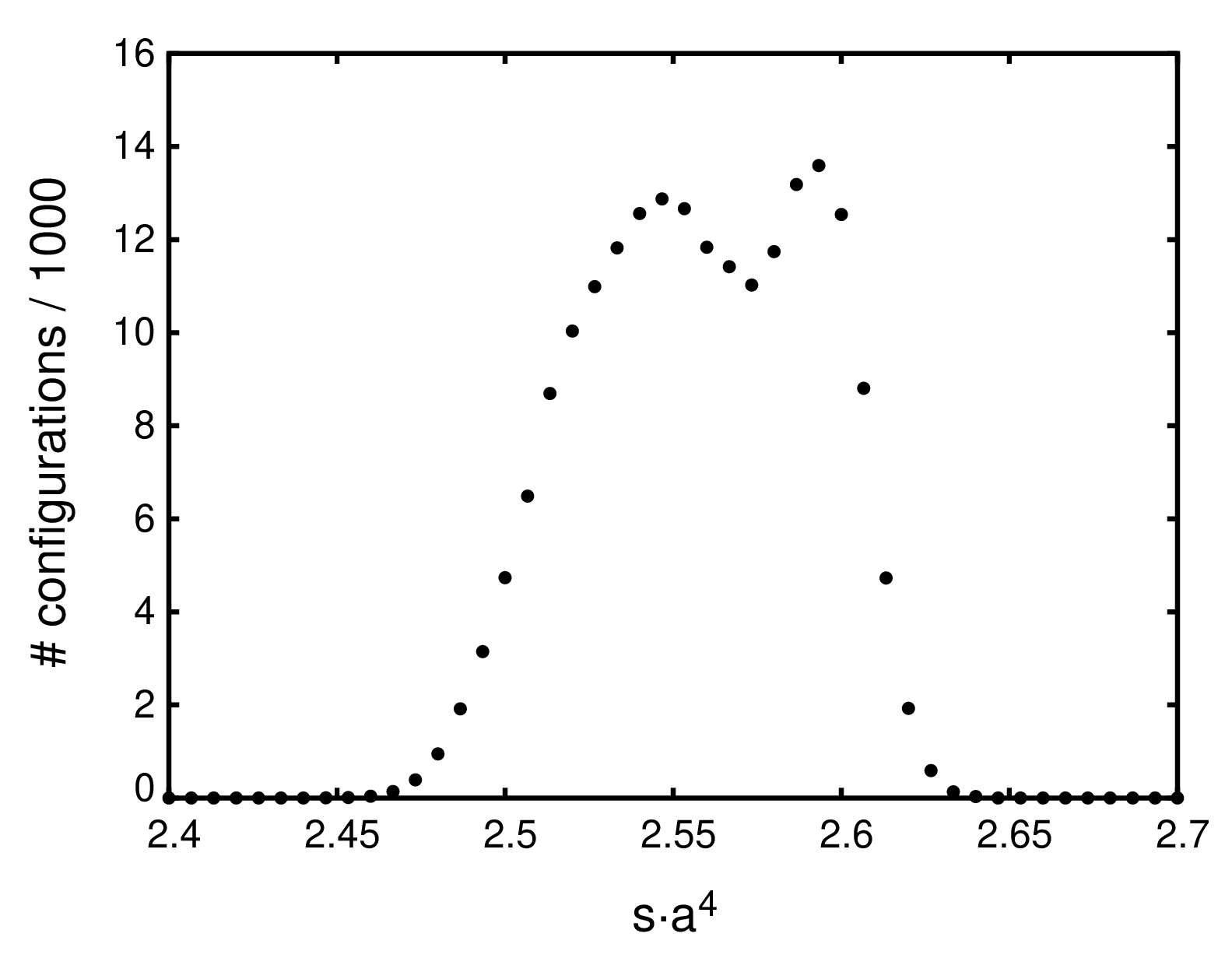,width=13cm} }
\caption{Distribution of the action density $s$ at the deconfining phase
transition, for an action containing a vortex world-surface curvature term
with $c_1 =0.5925$, $c_2 =0.7059$ and a branching term with $b=0.3800$. The
measurement was taken on a $20^3 \times 3$ lattice. Taking the distance
between the two maxima as a rough measure of the action density discontinuity
$\Delta s$ and supplementing this with a measurement of the zero-temperature
quark string tension $\sigma_{1} $ in lattice units at the same set of
coupling parameters, the condition (\ref{latin}) is satisfied by this
first-order transition.}
\label{bnt3}
\end{figure}

The physical point of the model was determined in analogy to the procedure
used in section~\ref{simpmod}. Properties of the deconfinement transition
generally are not directly accessible for the physical set of coupling
constants; instead, they are inferred by interpolation of phase transition
data obtained at different $N_t $. Specifically, for $N_t =1,2,3$ the
parameter sets were determined which realize the deconfinement temperature
while simultaneously satisfying both of the conditions (\ref{ratin}) and
(\ref{latin}). These parameter sets are listed in Table~\ref{physpt},
and corresponding action density distributions illustrating the
first-order behavior of the phase transitions are displayed in
Figs.~\ref{bnt1}-\ref{bnt3}.

\begin{table}[h]
\begin{center}
\begin{tabular}{|c||c|c|c||c|}
\hline
& $c_1 $ & $c_2 $ & $b$ & $T_c /\sqrt{\sigma_{1} } $ \\
\hline \hline
$N_t =1$ & 0.2785 & 0.4005 & 0.1403 & 0.90 \\
\hline
$N_t =2$ & 0.4558 & 0.7983 & 0.6950 & 0.61 \\
\hline
$N_t =3$ & 0.5925 & 0.7059 & 0.3800 & 0.50\\
\hline
\end{tabular}
\end{center}
\caption{Sets of coupling constants realizing the deconfinement temperature
as well as satisfying two of the three conditions defining the physical
point, namely (\ref{ratin}) and (\ref{latin}). The remaining condition
(\ref{tcds}) is subsequently satisfied by interpolating the parameters
using the data in the final column.}
\label{physpt}
\end{table}

The table furthermore records the ratio $T_c /\sqrt{\sigma_{1} } $
corresponding to each parameter set. To obtain the physical point, one
constructs the quadratic interpolations of the coupling constants $c_1 $,
$c_2 $ and $b$ as functions of $T_c /\sqrt{\sigma_{1} } $. Setting
$T_c /\sqrt{\sigma_{1} } $ equal to its physical value, cf.~(\ref{tcds}),
yields the physical point
\begin{equation}
c_1 = 0.45 \ \ \ \ \ \ \ c_2 = 0.80 \ \ \ \ \ \ \ b=0.71
\label{physparms}
\end{equation}
As is evident from Table \ref{physpt}, the physical point is very near
the parameters obtained for $N_t =2$; the uncertainties inherent in the
interpolation thus remain small. Since the conditions (\ref{ratin}) and
(\ref{latin}) were satisfied for all parameter sets used in the interpolation,
they are expected to be satisfied at the the physical point (\ref{physparms})
as well. In the case of (\ref{ratin}), this can be cross-checked. Measuring
$\sigma_{2} /\sigma_{1} $ directly at the physical point indeed yields
agreement with (\ref{ratin}). Of course, (\ref{latin}) cannot be
cross-checked directly at the physical point.

To appreciate the ultraviolet cutoff scale, it is useful to cast the
lattice spacing $a$ in physical units. Setting the zero-temperature quark
string tension to $\sigma_{1} = (440\, \mbox{MeV})^2 $, and combining this
with the measurement of that string tension in lattice units at the physical
point, $\sigma_{1} a^2 = 0.68$, yields
\begin{equation}
a = 0.37 \, \mbox{fm} \ .
\end{equation}
Within the framework of the random vortex world-surface model, this
simultaneously characterizes the transverse thickness of the vortices,
since it implements a minimal distance which parallel vortices must be
apart in order to remain distinguishable. In the present $SU(4)$ case,
this distance is slightly smaller than in the $SU(2)$ and $SU(3)$
cases, in which the minimal distance is $0.39 \, \mbox{fm} $. As
already noted further above, the fact that both of the two physically
distinct types of vortices present in the $SU(4)$ model are treated as
having the same thickness is a model restriction due to the present
hypercubic lattice formulation; in general, there is no physical
reason forcing the two thicknesses to be the same. Note also that a
decrease of the average vortex thickness as the number of colors is raised
seems plausible in view of the qualitative picture discussed in
section \ref{dyniss}, cf.~Fig.~\ref{monopfig}.

Having modeled the lattice Yang-Mills data presented in section
\ref{datasec}, it is possible to predict string tensions at finite
temperatures. In particular, the quantitative behavior of the
spatial string tensions $\sigma_{1}^{S} $ and $\sigma_{2}^{S} $ in the
deconfined phase can be accessed. At the physical point (\ref{physparms}),
a lattice with $N_t =1$ realizes the temperature\footnote{This temperature
determination is minimally adjusted compared to the value quoted in the
preliminary report \cite{su4proc} due to improved statistics in the
calculation of the zero-temperature string tension.} $T=1.95\, T_c $,
and one obtains
\begin{equation}
\sigma_{1}^{S} (T=1.95\, T_c ) / \sigma_{1} (T=0) = 1.34 \ \ \ \ \ \ \ \ \
\sigma_{2}^{S} (T=1.95\, T_c ) / \sigma_{2} (T=0) = 1.44 \ .
\end{equation}
While both string tensions exhibit the characteristic rise with temperature
in the deconfined phase, the behavior of $\sigma_{1}^{S} $ in particular is
quite different from the simplified model discussed in section \ref{simpmod}.
In contradistinction to that model, here, $\sigma_{1}^{S} $ rises less
strongly than $\sigma_{2}^{S} $, and the ratio of the spatial diquark string
tension to the spatial quark string tension is enhanced at $T=1.95\, T_c $,
\begin{equation}
\sigma_{2}^{S} (T=1.95\, T_c ) / \sigma_{1}^{S} (T=1.95\, T_c ) = 1.46
\label{ssratio}
\end{equation}
compared to the zero-temperature value $\sigma_{2} /\sigma_{1} =1.36$. This
should be contrasted with the behavior in the aforementioned simplified model,
where $\sigma_{2}^{S} (T=1.9\, T_c ) / \sigma_{1}^{S} (T=1.9\, T_c ) = 1.25$,
cf.~(\ref{simprat}).

Furthermore, in the more recent survey of $SU(4)$ lattice Yang-Mills
characteristics \cite{luctepnew}, which became available to the author
after the bulk of the numerical work in the present investigation was
completed, data for this string tension ratio are reported. According to
these data, the ratio $\sigma_{2}^{S} / \sigma_{1}^{S} $ remains near its
zero-temperature value in the temperature regime considered here. Thus,
while the comprehensive vortex model constructed in the present section
adjusts the value of $\sigma_{2}^{S} / \sigma_{1}^{S} $ in the correct
direction compared to the simplified model of section \ref{simpmod}, the
adjustment overshoots the $SU(4)$ lattice Yang-Mills result considerably.
The result (\ref{ssratio}) is almost as far above the lattice data as
the corresponding result in the simplified model, (\ref{simprat}), is
below those data.

It should be noted that readjusting the present model to reproduce updated
values from \cite{luctepnew} for the observables discussed in
section \ref{datasec} would of course impact this comparison. On the other
hand, regardless of the outcome of such an exercise, a further refinement
of the $SU(4)$ vortex model action would presumably permit fitting also the
finite-temperature spatial string tension ratio correctly, in addition to
the observables of section \ref{datasec}. For instance, one could contemplate
using a coupling constant $c_{12} \neq c_1 \cdot c_2 $ in the curvature
action (\ref{curvature}); while this option was discarded above due to its
physical similarity to the branching action (\ref{sbranch}), it is not
precisely the same and thus may allow for some additional quantitative
tuning. Another possible source of discrepancies between the model
constructed here and $SU(4)$ Yang-Mills theory is the artificial
restriction to a common thickness for both types of vortices present
for $SU(4)$ color. Refinements along these lines will not be considered
further here. If anything, such an endeavor would only further reinforce
the conclusions drawn in the next, concluding, section below.

\section{Conclusions}
The goal of this investigation was to construct a random vortex world-surface
model for $SU(4)$ Yang-Mills theory, particularly with a view towards the
question whether infrared $SU(4)$ Yang-Mills phenomenology forces one to
abandon the dynamical concept employed in the $SU(2)$ and $SU(3)$ cases,
which is based purely on vortex world-surface characteristics. General
arguments, put forward in \cite{jeffstef} and reviewed in section \ref{motiv},
suggest that a shift away from this type of dynamics occurs as the number
of colors is raised, with additional color structures present on the vortex
world-surfaces attaining their own dynamical significance and influencing
the vortex ensemble.

The modeling effort carried out in this work supports these arguments.
It proved impossible to construct a vortex model which reproduces the main
infrared characteristics of $SU(4)$ Yang-Mills theory based purely on a
vortex world-surface curvature action. Instead, an explicit weighting of
vortex branchings, which, depending on the color description, are 
associated with magnetic monopoles or nexi, turned out to furnish the
additional flexibility needed to match $SU(4)$ Yang-Mills theory.

The need to introduce new action terms of course impacts the predictive
capability of the model. Indeed, the effortless predictivity of the
$SU(2)$ and $SU(3)$ models, in which merely one dimensionless curvature
coefficient needed to be adjusted, is lost in the $SU(4)$ case. Three
dimensionless parameters were necessary to obtain a sufficiently flexible
model. After having determined these parameters, the behavior of the spatial
string tensions at high temperatures was predicted and compared to newer
lattice Yang-Mills data which became available subsequently. It should be
noted that the predictions are obtained at the ultraviolet limit of
validity of the vortex model, and thus can be expected to be more prone
to error than observables at lower temperatures; nevertheless, the
discrepancy in the ratio of the spatial diquark to the spatial quark string
tension at $T\approx 2T_c $, amounting to about 7\%, seems significant.
Aside from the possibility that this comparison may improve after
readjusting the present vortex model to reproduce updated values from
\cite{luctepnew} for the observables discussed in section \ref{datasec},
further model refinements, as suggested at the end of section \ref{bransec},
would presumably permit closing this gap in any case. However, this would
merely exacerbate the loss of predictive capability, already noted above,
compared with the $SU(2)$ and $SU(3)$ vortex models.

In view of this, it does not seem attractive to pursue the construction of
a yet more detailed $SU(4)$ vortex model. Indeed, the main question which
this investigation aimed to address can be answered in the affirmative
already at the present stage, as discussed above: As the number of colors
is raised beyond $N=3$ in $SU(N)$ Yang-Mills theory, the corresponding
infrared effective vortex description cannot continue to rely on dynamics
based purely on vortex world-surface characteristics. Solid evidence for
this emerges in the $N=4$ case studied in this work. To restore
the correct $SU(4)$ Yang-Mills phenomenology at the deconfining phase
transition, a generalization of the vortex model action was necessary, and
an explicit dependence of the dynamics on vortex branching characteristics
proved sufficient to achieve that goal.

\section*{Acknowledgments}
Fruitful discussions with M.~Quandt and H.~Reinhardt are acknowledged.
This work was supported by the U.S.~DOE under grant number DE-FG03-95ER40965.

\end{document}